\documentclass[5p,times,final]{elsarticle}

\usepackage{amssymb}
\usepackage{graphicx}
\usepackage{natbib}

\usepackage{url}
\usepackage{verbatim}
\usepackage{subfig}
\usepackage{epsfig}
\usepackage{graphicx}

\newcommand{\sectionname}{section}
\newcommand{\secref}[1]{\sectionname~\ref{#1}}
\newcommand{\figref}[1]{\figurename~\ref{#1}}

\begin{document}

\begin{frontmatter}

\title{Benchmarking mixed-mode PETSc performance on high-performance architectures}

\author[amcg]{Michael Lange\corref{cor1}}
\ead{michael.lange@imperial.ac.uk}
\author[amcg]{Gerard Gorman}
\ead{g.gorman@imperial.ac.uk}
\author[epcc]{Mich\`ele Weiland}
\author[epcc]{Lawrence Mitchell}
\author[stfc]{Xiaohu Guo} 
\author[fle]{James Southern}

\address[amcg]{Applied Modelling and Computation Group,
Imperial College London, London, UK}
\address[epcc]{EPCC, The University of Edinburgh, Edinburgh, UK}
\address[stfc]{Science and Technology Facilities Council,
Daresbury Laboratory, Warrington, UK}
\address[fle]{Fujitsu Laboratories of Europe Ltd., Hayes, Middlesex, UK}

\cortext[cor1]{Corresponding author}

\begin{abstract}

The trend towards highly parallel multi-processing is ubiquitous in
all modern computer architectures, ranging from handheld devices to
large-scale HPC systems; yet many applications are struggling to fully
utilise the multiple levels of parallelism exposed in modern
high-performance platforms. In order to realise the full potential of
recent hardware advances, a mixed-mode between shared-memory
programming techniques and inter-node message passing can be adopted
which provides high-levels of parallelism with minimal overheads. For
scientific applications this entails that not only the simulation code
itself, but the whole software stack needs to evolve.

In this paper, we evaluate the mixed-mode performance of PETSc, a
widely used scientific library for the scalable solution of partial
differential equations. We describe the addition of OpenMP threaded
functionality to the library, focusing on sparse matrix-vector
multiplication.  We highlight key challenges in achieving good
parallel performance, such as explicit communication overlap using
task-based parallelism, and show how to further improve performance by
explicitly load balancing threads within MPI processes.

Using a set of matrices extracted from Fluidity, a CFD application
code which uses the library as its linear solver engine, we then
benchmark the parallel performance of mixed-mode PETSc across multiple
nodes on several modern HPC architectures. We evaluate the parallel
scalability on Uniform Memory Access (UMA) systems, such as the
Fujitsu PRIMEHPC FX10 and IBM BlueGene/Q, as well as a Non-Uniform Memory
Access (NUMA) Cray XE6 platform. A detailed comparison is performed
which highlights the characteristics of each particular architecture,
before demonstrating efficient strong scalability of sparse
matrix-vector multiplication with significant speedups over the
pure-MPI mode.

\end{abstract}


\end{frontmatter}

\section{Introduction}
\label{sec:introduction}

Recent development in High Performance Computing (HPC) architectures
has been driven by a clear trend towards greater numbers of lower
power cores and a decreasing memory to core ratio. Numerical
algorithms and scientific software have to adapt to these changes to
efficiently utilise the available memory and network bandwidth.
Hybrid programming techniques, where shared memory programming is
combined with inter-node message passing, can be used to exploit the
multiple levels of parallelism inherent in modern architectures in
order to achieve sustainable scalability on massively parallel
systems.

In this paper we describe the addition of OpenMP thread parallelism to
the Portable Extensible Toolkit for Scientific Computation
(PETSc)~\cite{petsc-user-ref,petsc-efficient}. PETSc is a widely used
library for the scalable solution of partial differential equations
and is often used as a key component of large scientific applications.

Sparse matrix-vector multiplication (SpMV) is by far the most
computationally expensive component of sparse iterative linear
solvers~\cite{Schubert11}. Therefore we focus on optimising SpMV
within PETSc using hybrid programming techniques and evaluate strong
scaling performance on large numbers of compute nodes. We demonstrate
that using task-based parallelism to hide communication latency can
provide significant speedups over naive OpenMP
parallelisation. Further, explicit thread-level load balancing can be
used to gain an additional performance increase, resulting in
significantly improved scalability over pure-MPI implementations in
the strong scaling limit.

The matrices used for benchmarking our implementation are extracted
from the open source, general-purpose, multi-phase computational fluid
dynamics (CFD) code Fluidity~\cite{fluidity_manual_v4}. Fluidity
solves the Navier-Stokes equations and accompanying field equations on
arbitrarily unstructured finite-element meshes. It is used in areas
including geophysical fluid dynamics, computational fluid dynamics and
ocean modelling~\cite{Piggot08}.

\subsection{Sparse Matrix-Vector Multiplication}
\label{sec:matmult_parallel}

PETSc offers a wide range of high-level components required for linear
algebra, such as linear and non-linear solvers as well as
preconditioners. These are based on a suite of parallel data
structures which implement basic vector and matrix operations. The
most computationally expensive operation for solvers and
preconditioners alike is the multiplication of sparse matrices with an
input vector.

\begin{figure*}
 \centering 
 \subfloat[Diagonal matrix elements are multiplied with the
 local part of the vector while remote vector elements are gathered.]{ 
  \includegraphics[width=0.4\textwidth]{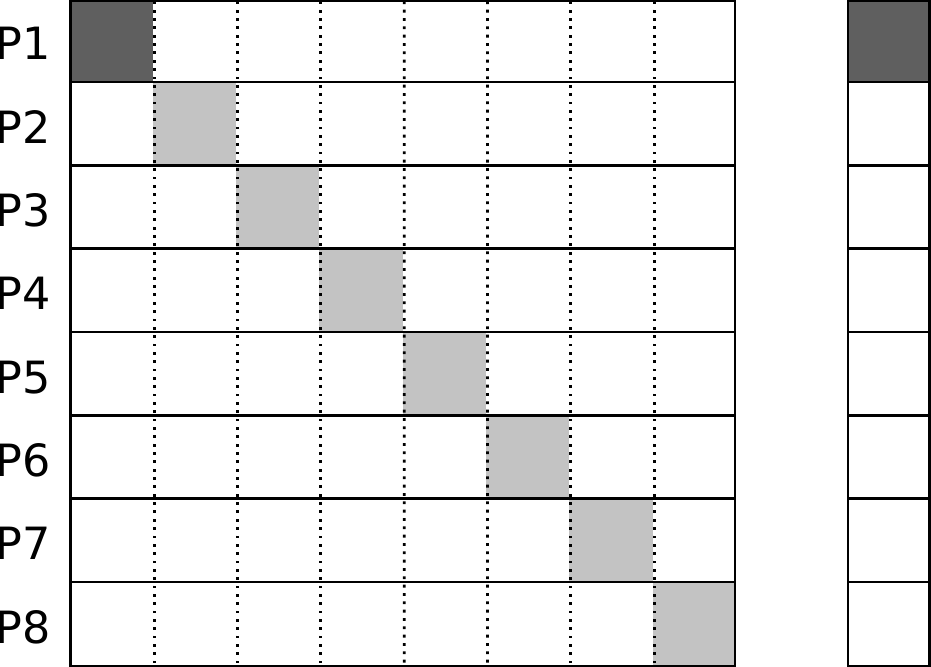}
  \label{fig:matmult_local}
 }
 \hspace{0.04\textwidth}
 \subfloat[The off-diagonal sub-matrix is then multiplied with a 
  local copy of the gathered vector elements and added to the partial solution.]{ 
  \includegraphics[width=0.4\textwidth]{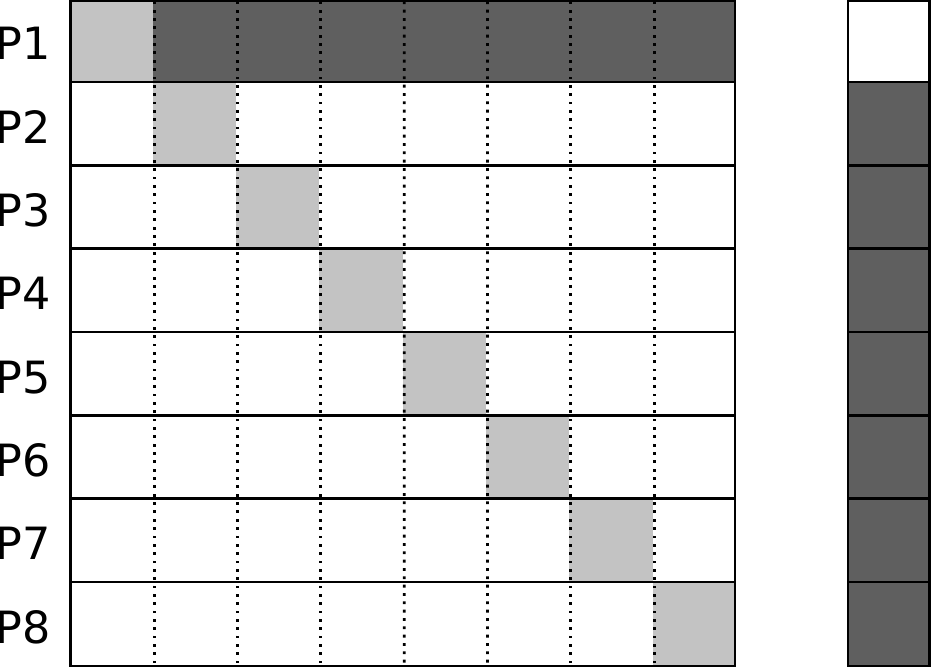}
  \label{fig:matmult_offdiag}
 }
 \caption{Parallel sparse matrix-vector multiplication using 8 MPI processes.}
 \label{fig:matmult_parallel}
\end{figure*}

PETSc represents distributed MPI matrices by dividing them into
diagonal and off-diagonal parts, which on each process are stored as
sequential matrices. The diagonal sub-matrix corresponds to
the part of the input vector that is stored locally by the process.
As a consequence of this storage strategy, as shown in
\figref{fig:matmult_parallel}, the matrix-vector multiplication is
implemented in two phases:
\begin{itemize}
\item First, each process multiplies its diagonal sub-matrix with the
 local elements of the input vector, while vector elements that
 reside off-process are gathered into the local memory of the
 executing process.
\item Off-diagonal matrix elements are then multiplied with the
 formerly remote vector elements and added to the partial solution.
\end{itemize}

\subsection{Related Work}
\label{sec:related_work}

Sparse matrix multiplication is one of the most heavily used kernels
in scientific computing and has therefore received attention from
several groups~\cite{Goumas2009,Williams2009,Bell2009,Rabenseifner2009}.
Multiple storage formats, optimisation strategies and even auto-tuning
frameworks exist to improve SpMV performance on a wide range of
multi-core architectures~\cite{Williams2009}. On modern HPC
architectures hybrid programming methods are being investigated to
better utilise the hierarchical hardware design by reducing
communication needs, memory consumption and improved load
balance~\cite{Rabenseifner2009}. In particular, task-based threading
methods have been highlighted by several researchers, where dedicated
threads can be used to overlap MPI communication with local work
~\cite{Wellein2003,Rabenseifner2009,Schubert11}.
 
\section{Hybrid MPI/OpenMP Parallelism}
\label{sec:hybrid}

Multi-core processors are now ubiquitous in HPC and programmers are
effectively presented with three levels of
parallelism~\cite{Robison10}:
\begin{itemize}
\item Between nodes, distributed memory parallelism is required to
  connect separate processors. This is most commonly implemented using
  explicit message passing via MPI.
\item Inside a compute node, cores share a contiguous memory address
  space and they can exchange information by directly manipulating
  this memory space.
\item Inside a core SIMD instructions may be applied to process
  multiple data items simultaneously, increasing the level of
  parallelism. For the purpose of this paper, however, we will not
  consider SIMD parallelism.
\end{itemize}
Exposing and expressing both intra- and inter-node parallelism can be
achieved using a hybrid programming approach, where message passing
between multiple compute nodes is complemented with thread-level
parallelism inside a node.

One motivation for moving away from MPI-only parallelised applications
is given by memory limitations. While the number of cores is steadily
increasing in modern HPC architectures, the memory available to each
core is decreasing~\cite{Rabenseifner2009}. By exploiting thread-level
parallelism, the same number of cores can be utilised within a single
node while reducing the MPI memory footprint~\cite{Balaji2009}. For
scientific applications based on domain decomposition, reducing the
MPI process granularity also reduces data replication due to halos or
ghost cells.

Performance gains may also be expected from using fewer MPI processes,
since it not only reduces communication overheads, but also improves
the load balance between individual
processes~\cite{Schubert11,Rabenseifner2009}. However, reducing
process-level imbalance may have a negative effect on the load balance
among threads, which in turn can be compensated for by node-level
scheduling strategies, as discussed in
\secref{sec:hybrid_load_balance}.

\subsection{NUMA Architecture}
\label{sec:hybrid_numa}

Non-Uniform Memory Access (NUMA) refers to multiprocessor systems
whose memory is divided into multiple memory nodes. This architecture
was designed to overcome the scalability limits of the symmetric
multiprocessing (SMP) architecture. However, this hierarchical memory
model for multi-core processors means that it takes longer for a
process or thread to access some parts of the memory than others.

It is therefore important to consider data locality in threaded
applications, since regular off-domain memory access can be
detrimental to the performance of already memory-bound applications.
In order to minimise bus contention a parallel \emph{first touch}
memory initialisation is often used on NUMA architectures to bind data
to the memory bank that is closest to the core subsequently using the
data block~\cite{Rabenseifner2009}. In addition, thread and process
pinning is required to optimise memory utilisation for all
bandwidth-bound algorithms.

When multiplying sparse matrices a master-only approach is most often
used to parallelise the local computation steps using threads (see
\secref{sec:matmult_parallel}). However, threaded SpMV across
multiple NUMA domains requires random but frequent off-domain memory
access to fetch input vector elements. In order to avoid the
high-latencies associated with off-domain data fetch NUMA domains can
be treated as single address spaces connected by multiple MPI tasks
within a compute node. This approach restricts threads to accessing a
single NUMA domain as demonstrated in \secref{sec:results_utilisation}.

\subsection{MPI-Communication Overlap}
\label{sec:hybrid_overlap}

As described in \secref{sec:matmult_parallel}, PETSc splits parallel
SpMV into two phases in order to allow the multiplication of the
diagonal submatrix to be overlapped with the MPI communication
required to fetch off-core vector elements. Nevertheless,
Schubert et al.~\cite{Schubert11} showed that few MPI implementations
provide truly asynchronous communication and significant performance
gains can be achieved by using \emph{task-based} threading, where a
single thread is dedicated to actively perform the localisation of
global vector elements. This approach not only overlaps MPI transfer
latencies with computation but also hides any sequential overhead
incurred from moving data to and from the required MPI buffer space.

\emph{Task-based} threading stands in contrast to traditional
\emph{vector-based} threading, where all threads share the
computational load evenly. In order to utilise the \emph{task-based}
variant the thread-parallel section needs to be lifted to enclose the
vector scatter-gather operation to localise input vector
elements. This prohibits the use of OpenMP \verb|parallel for| pragmas
to distribute the local row-wise computation among threads and
requires the explicit computation of thread partition boundaries.

\subsection{Thread-level Load Balance}
\label{sec:hybrid_load_balance}

Traditional \emph{vector-based} threading with OpenMP divides the
number of matrix rows approximately evenly among threads by applying
\verb|parallel for| pragmas to the outer loop. This, however, ignores
the fact that individual rows may incur varying amounts of
computational work, creating a potential load imbalance within
individual thread groups. Instead, thread-level load balance may be
improved statically by dividing the number of non-zeros approximately
equally between threads, as pointed out by Williams et
al.~\cite{Williams2009}.

It is important to note that the matrix stencil does not change during
the solve. Thus, an explicit thread partitioning scheme may be
computed after the matrix has been assembled and cached with the
matrix object. This turns the load balance optimisation into a
one-off cost, allowing, in principle, the use of load balancing
schemes of arbitrary complexity.

The method used in this paper starts with an initial greedy
allocation, where each worker thread receives a block of continuous
rows. This is followed by an iterative local diffusion algorithm,
which further balances the number of non-zeros allocated to each
thread. This procedure balances the thread-level work load and memory
bandwidth requirement according to floating point operations required
for the solution.

\section{Benchmark}
\label{sec:benchmark}

The matrices used for benchmarking the hybrid MPI/OpenMP
implementations have been generated by Fluidity from a global
baroclinic ocean simulation, which is representative of a range of
three-dimensional multi-scale oceanographic problems~\cite{Piggot08}.
The unstructured mesh is based on two-dimensional high-resolution
coastline data that is extruded vertically using constant spacing. By
changing the vertical resolution of the extruded mesh the size of the
problem can be scaled linearly, allowing a controlled quasi-linear
increase in work load for the extracted matrices.

The benchmark matrices used in this work are pressure field solves
extracted after five timesteps. The resulting matrices are solved
using the Conjugate Gradient method with a Jacobi preconditioner and
the number of iterations was limited to $10,000$.

\subsection{Cray XE6}
\label{sec:hector}

One of the benchmarking systems used for the work presented here is
HECToR, a Cray XE6 based on the AMD Opteron 6200 Interlagos processor
series and Crays Gemini interconnect~\cite{cray_xe6}. The Interlagos
compute nodes are based on two AMD Bulldozer processors, each with 16
cores at $2.3$ GHz paired into two modules providing a theoretical
peak performance of $163.7$ GFlop/s per module and $327.4$ GFlop/s per
node. Each module has its own associated memory bank, which provides
a peak memory bandwidth of $51.2$ GB/s, resulting in four separate
memory nodes per compute node~\cite{bulldozer}.

\subsection{Fujitsu PRIMEHPC FX10}
\label{sec:fx10}

The second benchmarking system available to us is a 96-node Fujitsu
PRIMEHPC FX10 system. The PRIMEHPC FX10 is a UMA
(Uniform Memory Access) architecture based on the SPARC64 IXfx
processor. A single compute node has 16 cores at $1.848$ GHz and a
peak memory bandwidth of $85$ GB/s, providing a theoretical peak
performance of $236.5$ GFlop/s~\cite{fujitsu_fx10}.

\subsection{IBM BlueGene/Q}
\label{sec:bgq}

The third system benchmarked in this paper is an IBM BlueGene/Q, a UMA
architecture based on the PowerPC A2 processor. Each node provides
access to 16 cores running at $1.6$ GHz and a memory bandwidth of
$42.6$ GB/s, providing a theoretical peak performance of $204.8$
GFlop/s~\cite{bgq_redbook}. Each PowerPC A2 core also provides
4-way simultaneous multi-threading (SMT) in hardware, where each core
executes up to four threads with little context switching overhead.

\section{Results: Hardware Utilisation}
\label{sec:results_utilisation}

Hybrid programming offers a complex set of choices on how to best
utilise a given hardware set. We therefore start our investigation by
analysing various process-to-thread ratios for fixed numbers of
cores, focusing on specific hardware features such as UMA/NUMA memory
architectures as well as simultaneous multi-threading (SMT) on the
BlueGene/Q architecture. This provides insights into the resource
utilisation of each algorithm and provides an estimate for the best
hybrid configuration to be used during the subsequent strong
scalability study on large numbers of compute nodes.

\begin{figure}
\centering
\subfloat[128 cores]{ 
\includegraphics[width=0.5\textwidth]{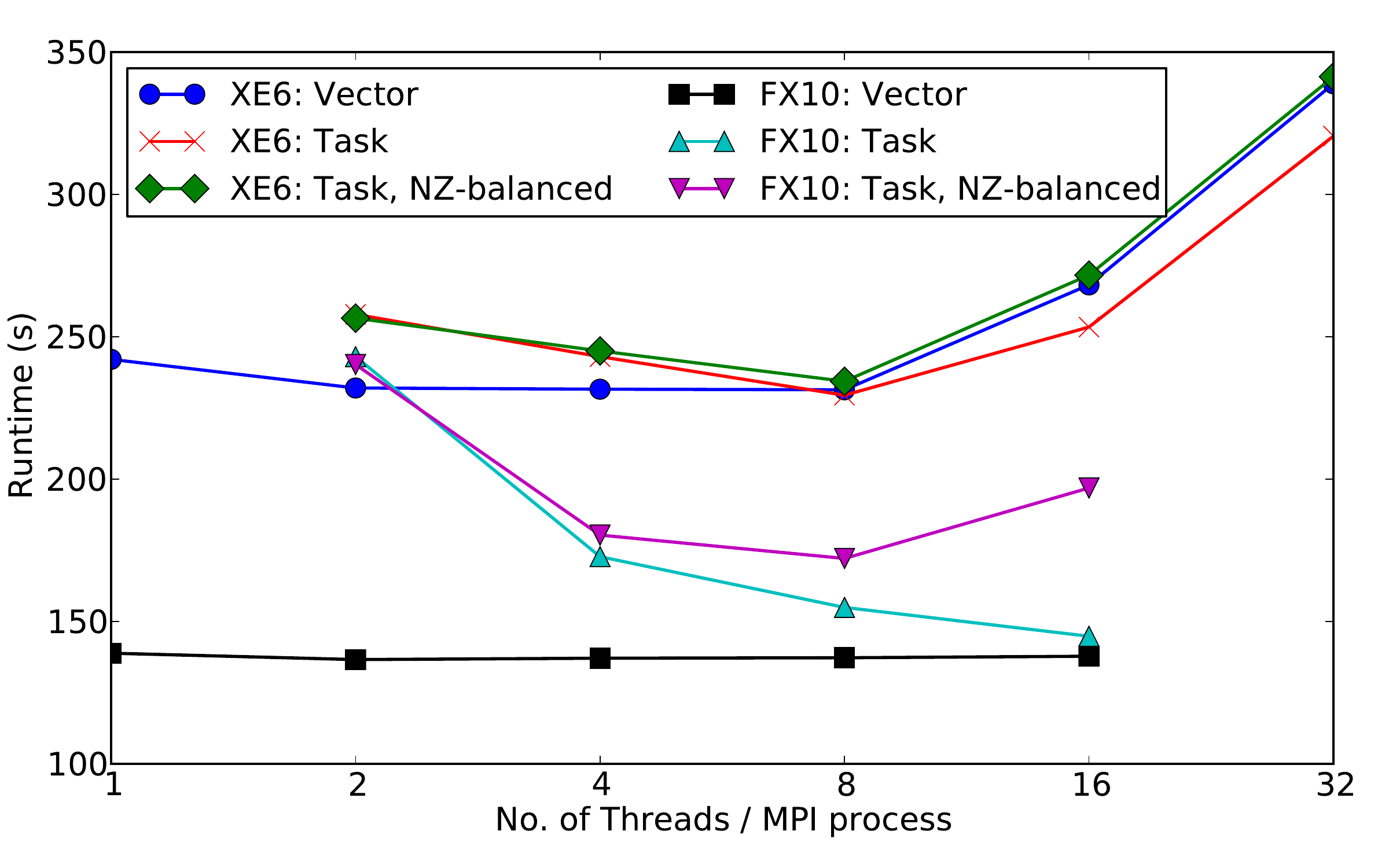}
\label{fig:sweetspot_nc128}
}\\
\subfloat[1024 cores]{ 
\includegraphics[width=0.5\textwidth]{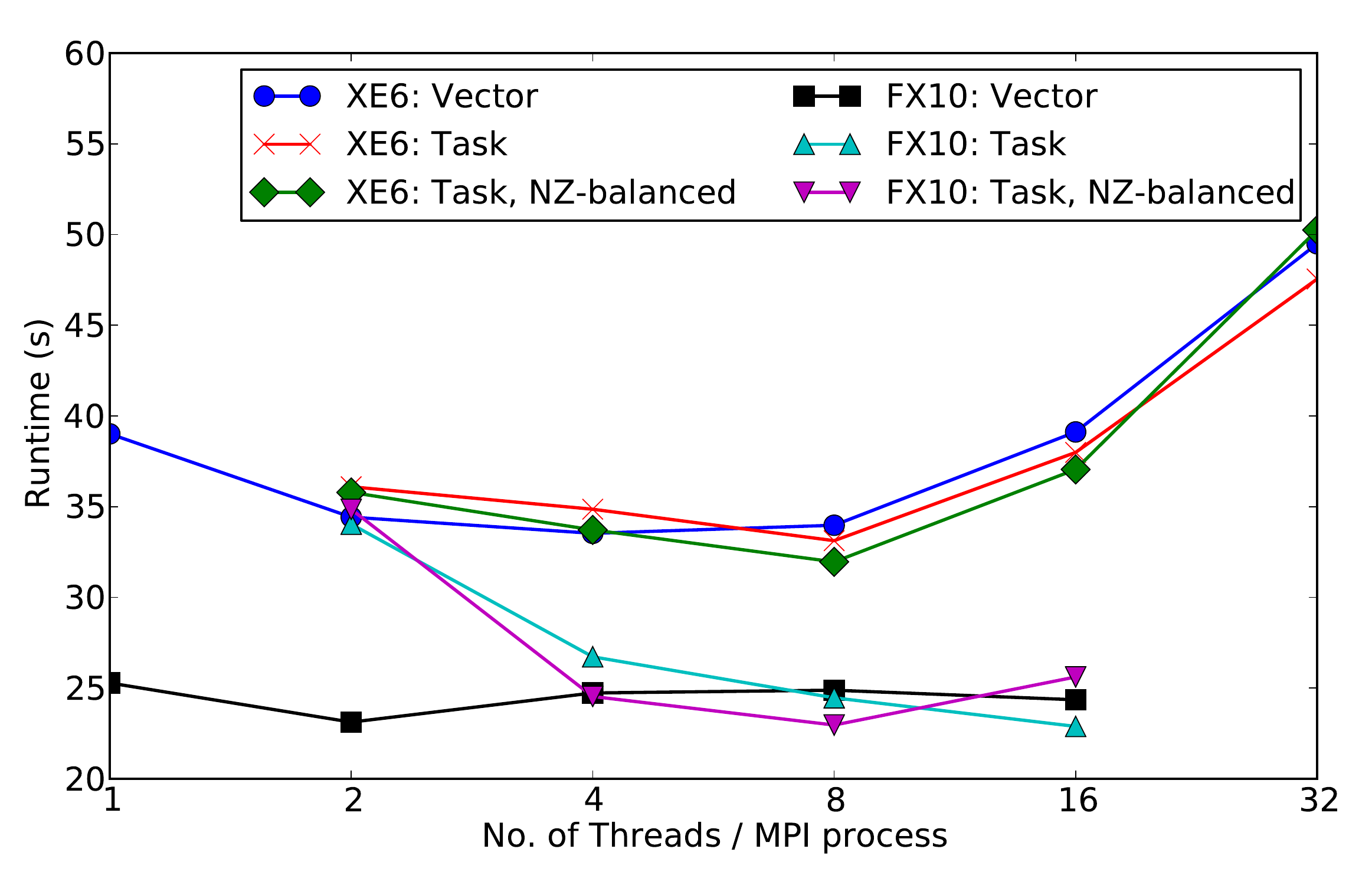}
\label{fig:sweetspot_nc1024}
}\\
\subfloat[4096 cores]{
\includegraphics[width=0.5\textwidth]{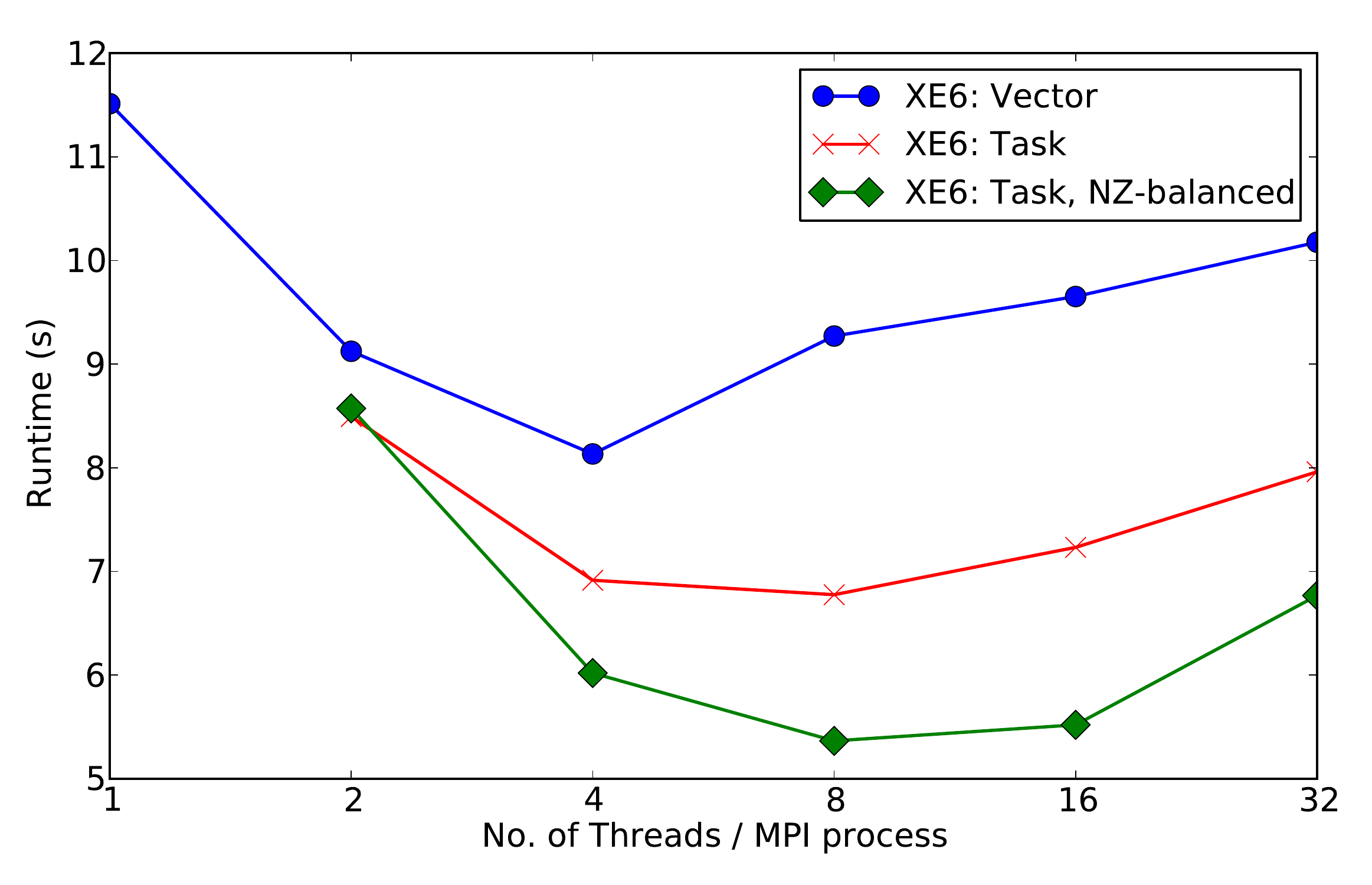}
\label{fig:sweetspot_nc4096}
}
\caption{
Matrix multiplication run times on a fixed number of cores with
varying thread-to-process ratios on Cray XE6 and Fujitsu PRIMEHPC
FX10. The left most value represents a close approximation to
MPI-only performance. Native compilers were used on both
architectures.
}
\label{fig:sweetspot}
\end{figure}

\figref{fig:sweetspot} shows the performance of varying
hybrid process-thread combinations on the Cray XE6 and Fujitsu
PRIMEHPC FX10 systems. The left-most entry of the vector-based
configuration constitutes the MPI-only baseline configuration. OpenMP
overheads have been verified to be negligible for the given problem
size using microbenchmarks~\cite{Reid2004}.

\subsection{Cray XE6}
\label{sec:results_utilisation_xe6}

On the Cray XE6 the task-based algorithms with and without explicit
thread-level load balancing perform best when running 8 threads
wrapped by 4 MPI processes per node. This correlates with NUMA
alignment, where threads are used only inside individual NUMA domains
and MPI tasks connect separate memory nodes. A significant performance
reduction can then be observed with 16 and 32 threads per process,
which coincides with NUMA traffic being incurred due to fetching input
vector elements (see \secref{sec:hybrid_numa}).

However, with an increasing number of cores, as highlighted with 4096
cores in \figref{fig:sweetspot_nc4096}, the performance penalty
incurred due to NUMA traffic with 16 and 32 threads per process is
less severe for all threading models. We can conclude that
the algorithm is now bound by memory bandwidth rather than
latency. Moreover the performance penalty incurred by going from 16 to
32 threads per process using the explicit thread balancing method is
greater due to the computed thread allocation contradicting the
original {\em first touch} memory allocation and therefore causing
additional NUMA traffic.

Furthermore, both task-based modes significantly outperform the
vector-based threading approach on 4096 cores, demonstrating the
performance loss due to MPI communication overheads. Although
vector-based threading provides better memory bandwidth utilisation on
small numbers of cores due to having an extra worker thread, on large
numbers of compute nodes the approach struggles to utilise the given
memory bandwidth, as shown in \figref{fig:sweetspot_nc4096}. Here the
performance is greatest with only four threads per process, indicating
that the algorithm's performance is bound by an unmasked communication
overhead that increases with the number of MPI processes.

\subsection{PRIMEHPC FX10}
\label{sec:results_utilisation_fx10}

On the PRIMEHPC FX10 system, we observe similar scaling properties and
resource limitations with an increasing number of processing cores for
all three algorithms. Thus, although the test system used for this
work was limited to 1536 cores, we can infer an estimate of the
scalability characteristics of the PRIMEHPC FX10 architecture for
large scale systems.

The key difference to the XE6 is that PRIMEHPC FX10 is a UMA
architecture, and therefore does not incur memory latency penalties
due to using multiple memory nodes per thread group. This can be
observed in \figref{fig:sweetspot_nc128} and
\ref{fig:sweetspot_nc1024} where, in contrast to the XE6, the
task-based mode without thread balancing improves performance steadily
with increasing numbers of threads per node. However, the explicit
thread balancing method experiences a slight slowdown using the
maximum number of threads per node due to an imbalance in vector
elements required by each thread aggravating the algorithm's sensitivity to
memory latency.

\begin{figure}
\centering
\subfloat[1024 cores]{ 
\includegraphics[width=0.5\textwidth]{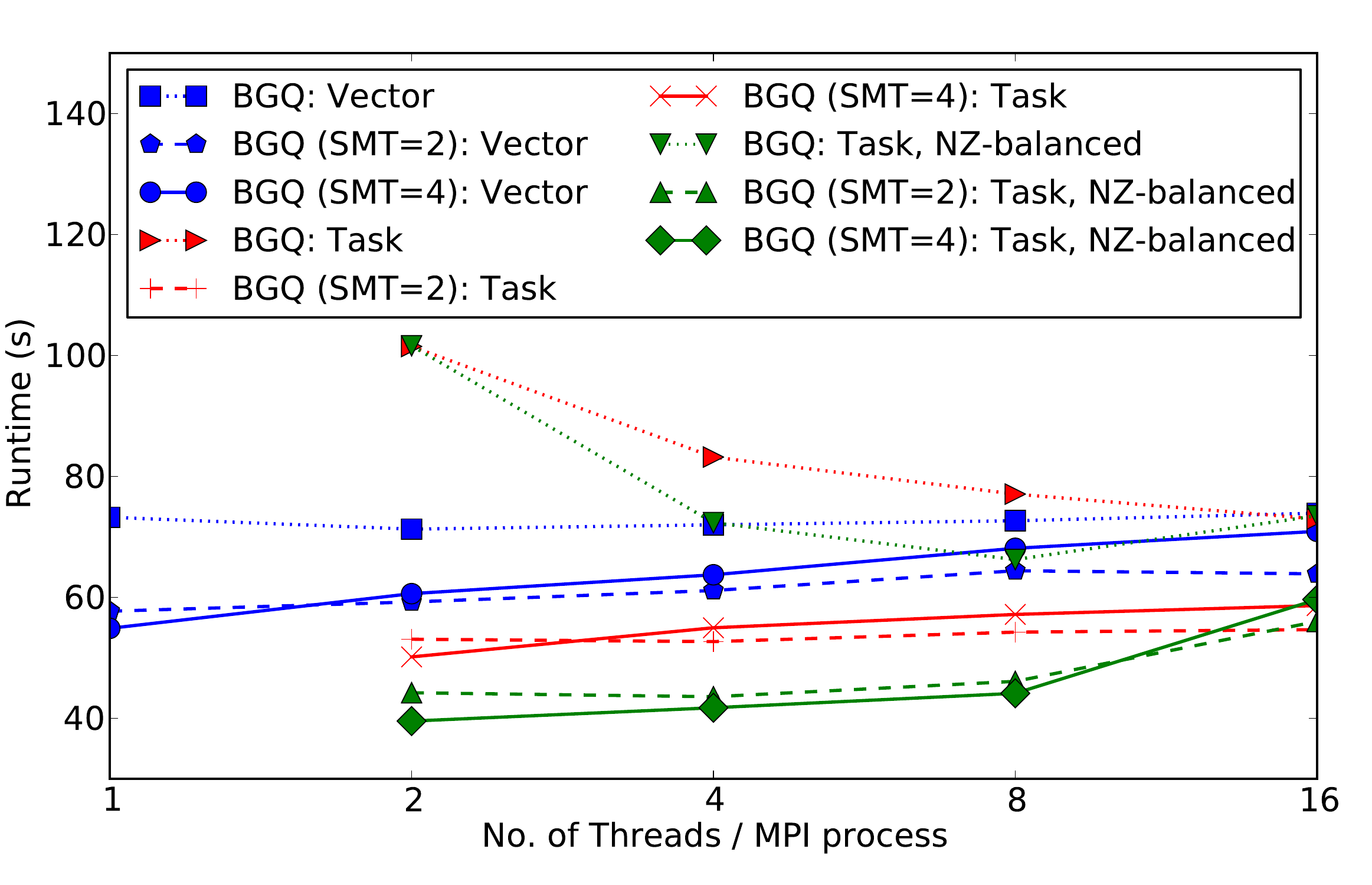}
\label{fig:sweetspot_bgq_1024}
}\\
\subfloat[4096 cores]{ 
\includegraphics[width=0.5\textwidth]{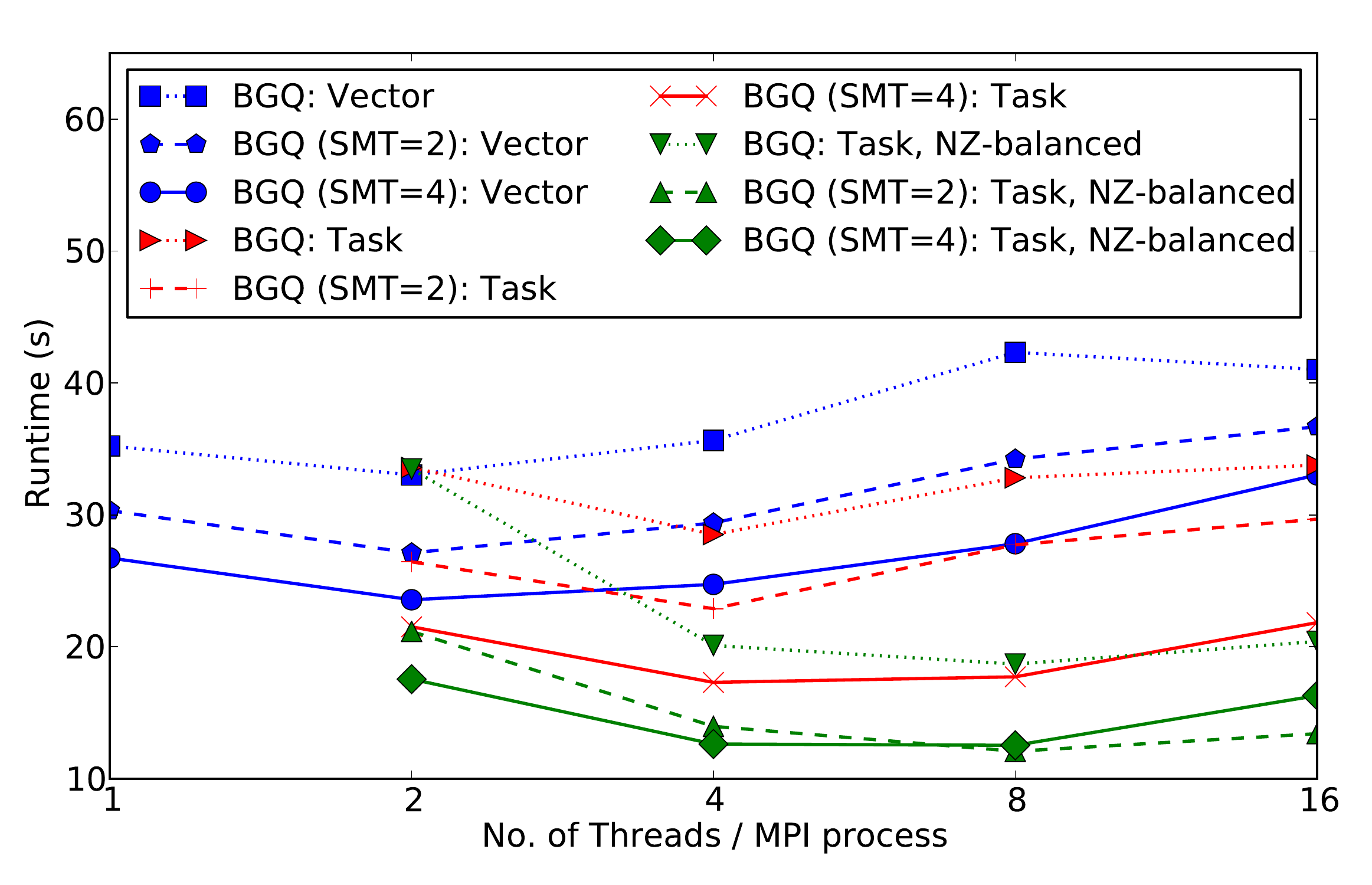}
\label{fig:sweetspot_bgq_4096}
}
\caption{
Matrix multiplication performance with varying thread-to-process
ratios and SMT depths on IBM BlueGene/Q. The number of
threads-per-process is multiplied with the SMT depth to align the
plots.
}
\label{fig:sweetspot_bgq}
\end{figure}

On 1024 cores (64 PRIMEHPC FX10 nodes,
\figref{fig:sweetspot_nc1024}), the profiles exhibit properties
similar to the 4096-core XE6 results. The vector-based mode is
limited by inter-process communication and performs best with two
threads per process, while the overall best performance is achieved by
the thread-balancing approach using eight threads per process.

\subsection{Multi-Threading on BlueGene/Q}
\label{sec:results_utilisation_bgq}

When analysing the performance of the different threading models with
varying thread-to-core ratios on the BlueGene/Q architecture we also
have to consider the number of threads running on each core (SMT
depth). In \figref{fig:sweetspot_bgq} the number of threads per process
is therefore multiplied with the SMT depth to align the plots, so
that, for example, a run with a thread-to-process ratio of 16 and
SMT depth of four uses 64 threads per BlueGene/Q node. A breadth-first thread
layout is hereby used to avoid unused cores with under-utilised node
configurations.

Using 1024 cores (64 BlueGene/Q nodes,
\figref{fig:sweetspot_bgq_1024}) all threading models exhibit a clear
speedup when using more than one thread per core. The speedup is
particularly strong with small thread-to-process ratios, indicating
that the algorithm is indeed bound by memory latency which can be
masked by increasing the SMT depth. This furthermore highlights the
importance of utilising the SMT feature of the BlueGene/Q hardware.

The difference in performance between four threads per core (fully
populated nodes) and two threads per core (half populated nodes) on
the other hand is not as clear. On 1024 cores
(\figref{fig:sweetspot_bgq_1024}) all threading models perform
slightly better with increasing nodes when using half populated nodes.
However, using 4096 cores (256 BlueGene/Q nodes,
\figref{fig:sweetspot_bgq_4096}) the vector-based approach and the
task-balancing method without explicit load balancing perform
constantly better when using four threads per core, whereas the
load-balanced implementation shows a growing performance advantage
when using half populated nodes with an increasing thread-to-process
ratio. The overall best performance is achieved with the explicit
thread-balancing scheme and a thread-to-process ration of eight on a
half populated hybrid node configuration.


\section{Results: Strong Scaling}
\label{sec:results_scaling}

In this section we analyse the strong scalability of the described
hybrid algorithms on the Cray XE6 and IBM BlueGene/Q system and
compare their performance to a pure-MPI approach. On the Cray XE6 all
hybrid modes were run using four MPI processes per compute node with
eight threads each in order to prevent NUMA traffic as described in
\secref{sec:results_utilisation_xe6}.

\begin{figure*}
\centering
\subfloat{
\includegraphics[width=0.9\textwidth]{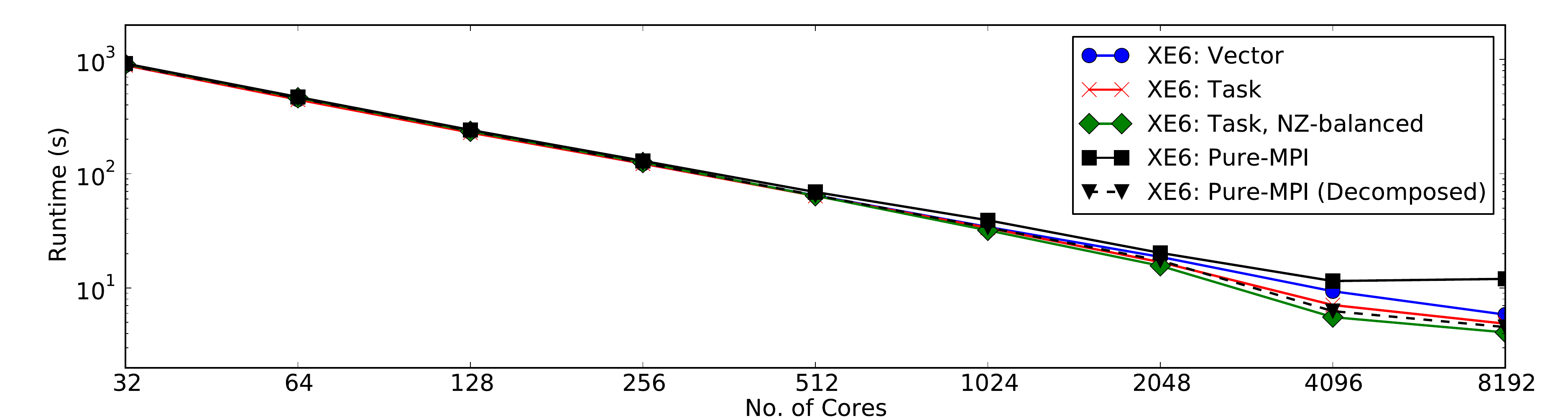}
\label{fig:results_1K_rt}
} \\
\subfloat{ 
\includegraphics[width=0.9\textwidth]{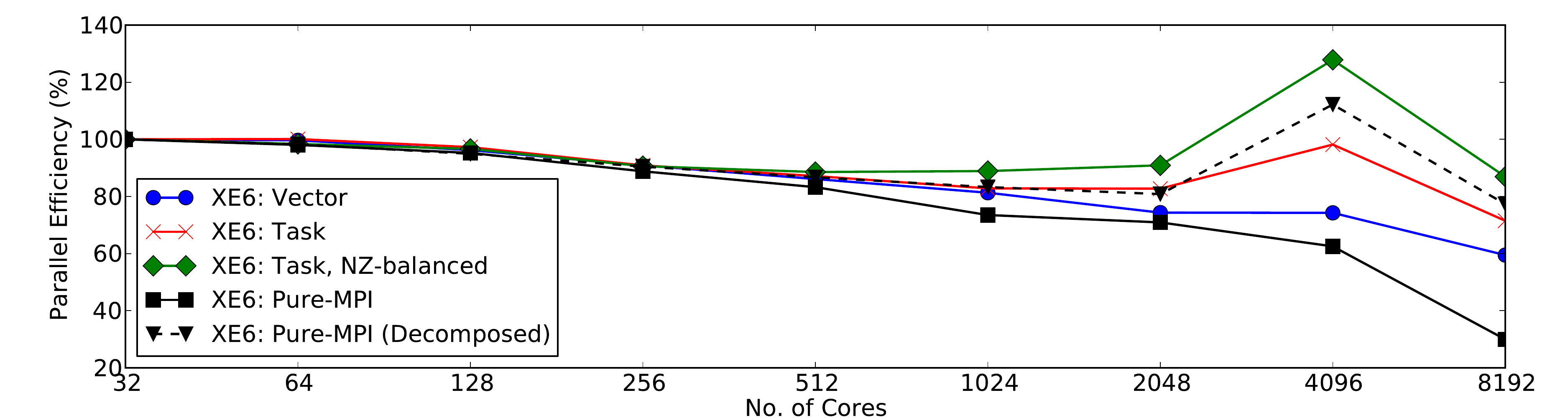}
\label{fig:results_1K_eff}
}
\caption{
Strong scaling results for the pressure matrix on up to 256 XE6 nodes
(8192 cores). All hybrid modes use 4 MPI ranks per node and 8 threads
per rank.
}
\label{fig:results_1K}
\end{figure*}

The matrix used in \figref{fig:results_1K} has a dimension of 13,491,933
rows and columns and 371,102,769 non-zero elements and was
generated by a parallel Fluidity simulation decomposed into 1024
sub-domains. For the hybrid modes the number of MPI processes used in
the strong limit therefore matches the number of processes used during
the original decomposition. For more than 1024 cores, however, the
pure-MPI mode uses more processes than the matrix was originally
optimised for, resulting in a potential slowdown due to load
imbalance. Therefore, an equivalent matrix which has been optimised
for 8192 MPI processes has also been included in the benchmark (dashed
line).

\begin{figure*}
\centering
\subfloat{
\includegraphics[width=0.9\textwidth]{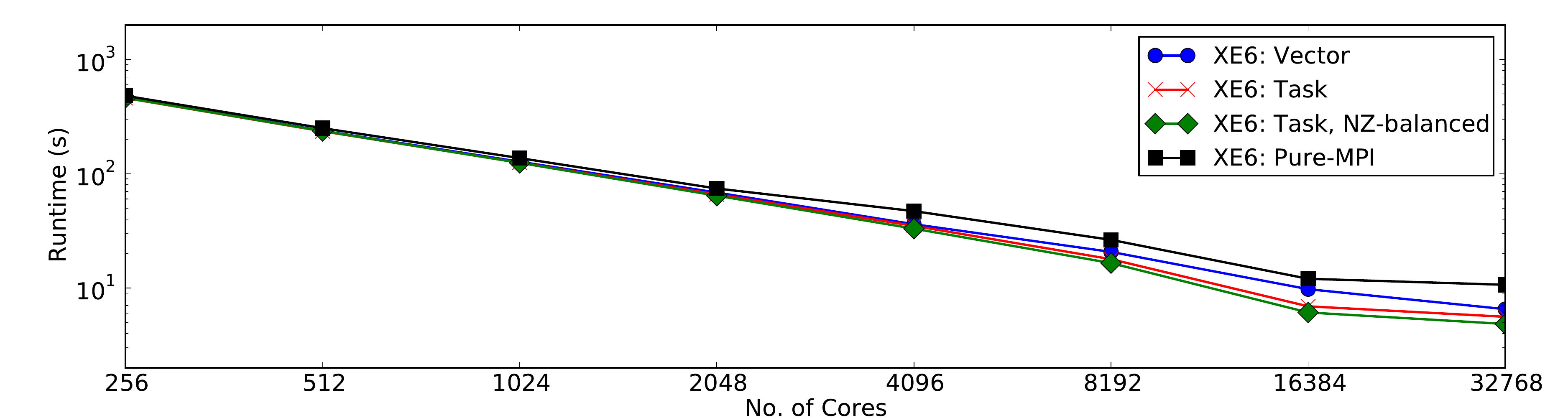}
\label{fig:results_x4_rt}
} \\
\subfloat{ 
\includegraphics[width=0.9\textwidth]{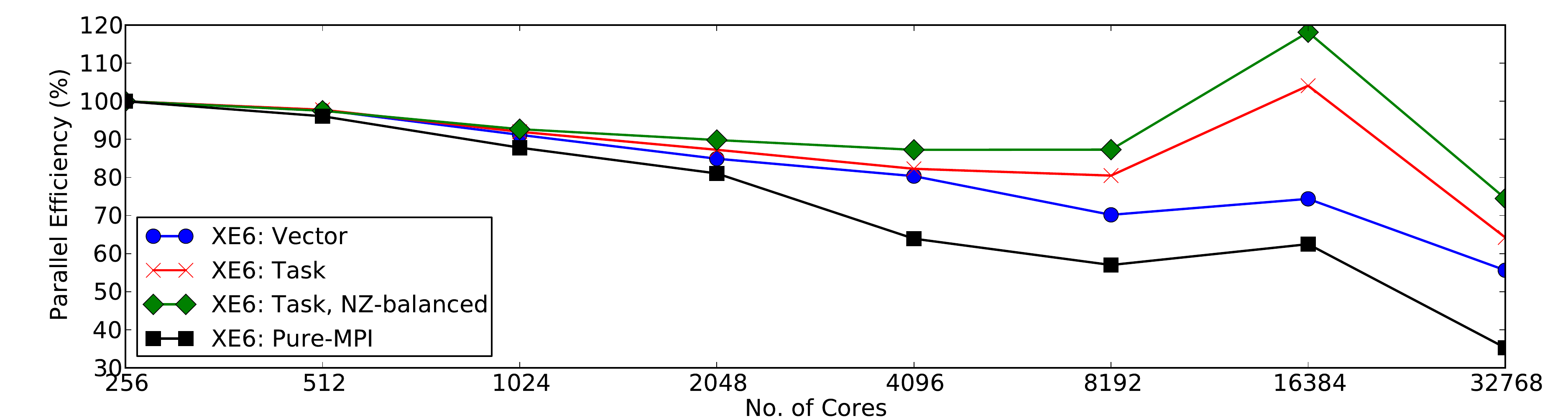}
\label{fig:results_x4_eff}
}
\caption{
Strong scaling results for a larger pressure matrix on up to 
1024 XE6 nodes (32768 cores). All hybrid modes use 4 MPI ranks per node 
and 8 threads per rank. Runs with less than 256 cores (8 XE6 nodes) 
have been omitted due to insufficient memory per MPI process.
}
\label{fig:results_x4}
\end{figure*}

\subsection{Cray XE6}
\label{sec:results_scaling_xe6}

At the low end of the scaling curve no significant performance
differences can be noted. For more than 512 cores (16 XE6 nodes) the
task-based hybrid methods show a better scalability over the
vector-based approach. The thread-balancing implementation
performs best, maintaining a nearly constant parallel efficiency of
$>88\%$ between 512 and 2048 cores, and even experiences slightly
super-linear scaling between 1024 and 2048 cores.

On the same matrix, the pure-MPI performance decreases significantly
faster than the hybrid algorithms for more than 512 cores (16 XE6
nodes). The equivalent MPI runs using a more finely decomposed
matrix, on the other hand, closely match the performance of the
task-based mode without thread-balancing up to 2048 cores. However,
in the strong limit the thread-balancing mode outperforms the
optimised MPI runs.

Furthermore, between 2048 and 4096 cores (64 and 128 XE6 nodes) we
observe strong super-linear scaling for both task-based methods.
Since the final runtime in the strong limit is below $4$ seconds, we
can deduce that scalability ceases at this point due to a lack of
computational work and that the super-linear scaling effects are due
to favourable cache effects.

\figref{fig:results_x4} shows scalability on up to 32,768 cores (1024
XE6 nodes) when the workload of the matrix multiplication is increased
by a factor of 4 by changing the vertical extrusion of the parent mesh
(see \secref{sec:benchmark}). This matrix has a dimension of
52,040,313 rows and columns and 1,462,610,289 non-zero elements and is based
on a 4096-domain partitioning. The results follow the same general
trend, with significant differences in performance observable in the
strong end of the scalability curve. The pure-MPI performance starts
to deteriorate earlier and the super-linear scaling in the high end is
more pronounced for all approaches.

\subsection{BlueGene/Q}
\label{sec:results_scaling_bgq}

\begin{figure*}
\centering
\subfloat{
\includegraphics[width=0.9\textwidth]{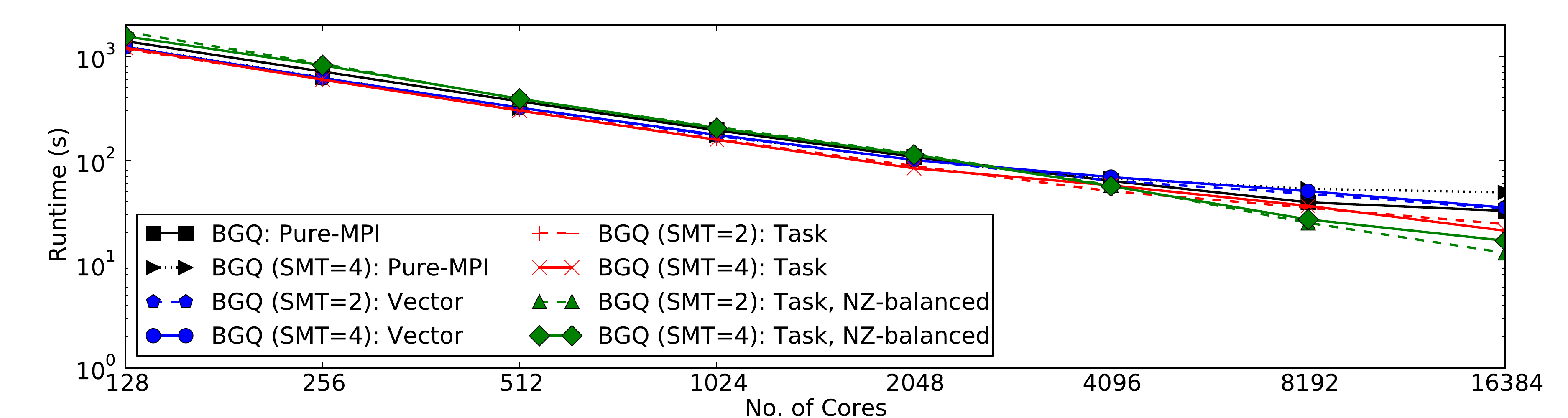}
\label{fig:results_bgq_x4_rt}
} \\
\subfloat{ 
\includegraphics[width=0.9\textwidth]{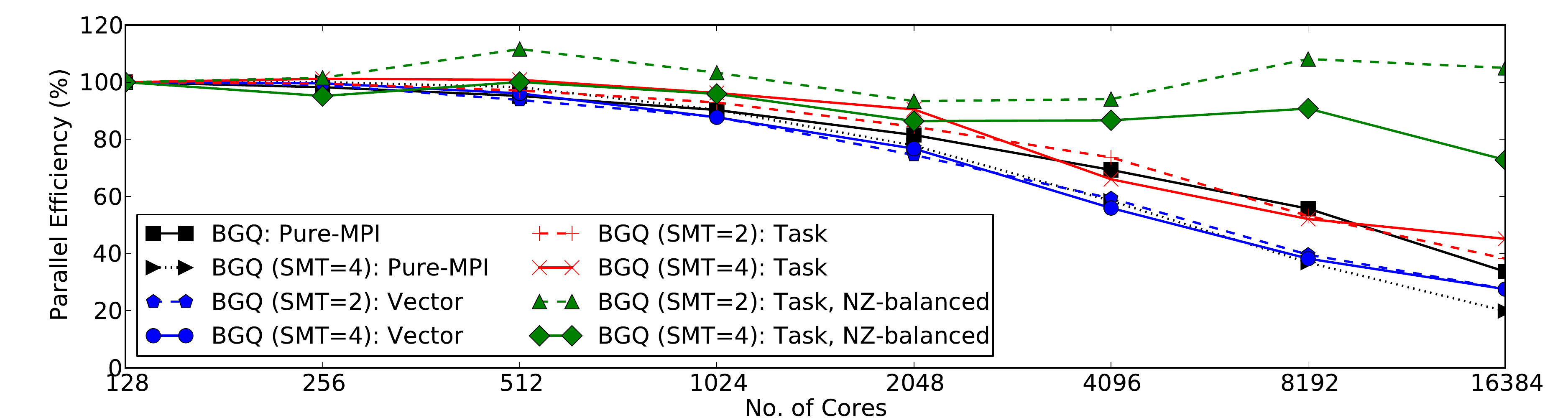}
\label{fig:results_bgq_x4_eff}
}
\caption{
Strong scaling results for the large pressure matrix on up to 1024
BlueGene/Q nodes. All hybrid modes use a single MPI process and 32
threads (SMT$=2$) or 64 threads (SMT$=4$) per node. The MPI (SMT$=4$)
runs use 64 MPI ranks per node.
}
\label{fig:results_bgq_x4}
\end{figure*}

The large version of the matrix used in \figref{fig:results_x4} has
also been used to evaluate the scalability of the varying approaches
on the BlueGene/Q architecture. Since the performance evaluation
performed in \secref{sec:results_utilisation_bgq} did not highlight an
optimal SMT configuration the scaling runs have been performed with
two and four threads per core. In addition to pure-MPI performance
using a single rank per core an equivalent set of runs was performed
which uses 64 MPI ranks per node to fully saturate all available hardware
threads.

In the low end of the scaling curve, up to around 1024 cores, the
vector-based approach and the task-based approach without
load-balancing provide the fastest runtime on the BlueGene/Q. However,
with increasing numbers of cores the task-based approach with
explicit thread-balancing maintains high scalability, whereas all
other approaches suffer from a steadily decreasing parallel
efficiency. 

Moreover, the scalability of the thread-balancing method in the high
end is improved even further by using only half saturated cores. For
more than 8192 cores the super-linear scaling effects observed in
\figref{fig:results_x4} are also achieved, in addition to super-linear
scaling in the low end. An overall parallel efficiency of $>90\%$ is
maintained throughout, clearly outperforming all other approaches on
16384 cores. The performance improvement from using only two threads
per core stems from reduced cache thrashing, where competing threads
cause the premature eviction of data from the shared L1 cache due to
the low computations-to-data-access ratio of the SpMV algorithm.

\section{Architecture Comparison}
\label{sec:arch_comparison}

\begin{figure}
\centering
\includegraphics[width=0.5\textwidth]{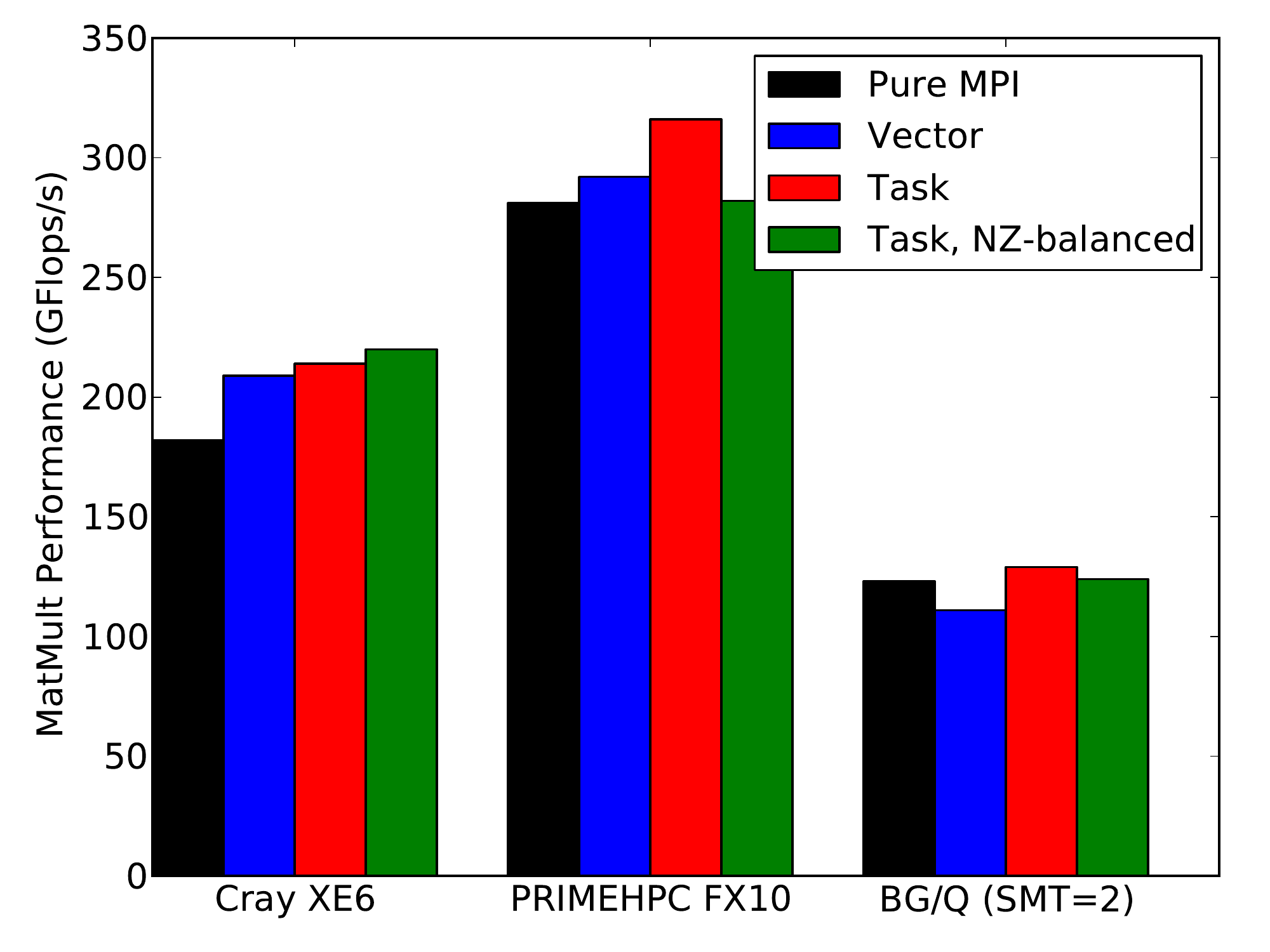}
\caption{
Comparison of achieved performance in GFlop/s on 1024 cores across all
architectures. The Cray XE6 runs use 8 threads per MPI process and
the BlueGene/Q results use two threads per core.
}
\label{fig:perf_comp}
\end{figure}

In this section we provide a comparison of the achieved performance on
all three architectures under consideration in this paper. In
\figref{fig:perf_comp} the achieved performance is shown for each
threading model on 1024 cores of each benchmark architecture using the
benchmark matrix described in \secref{sec:results_scaling}. All hybrid
runs were aligned with the memory domains, using a single MPI instance
on both UMA architectures and four MPI ranks on the Cray XE6.

The most significant performance improvement of the optimised
task-based threading models over a pure MPI implementation can be
observed on the Cray XE6. The PRIMEHPC FX10 also shows improved
performance when using the task-based variant without explicit
thread-balancing. However, it should be kept in mind that, as shown
in \figref{fig:sweetspot_nc1024}, a similar performance can be
achieved with the thread-balancing option when using fewer threads per
MPI process. The same is true for the BlueGene/Q architecture, where,
although no clear advantage from using the advanced threading models
can be seen in \figref{fig:perf_comp}, \figref{fig:sweetspot_bgq_1024}
shows that an even larger percentage of the peak performance is
attainable with less threads per MPI rank.

Overall the Fujitsu PRIMEHPC FX10 obtains highest performance. This is
largely due to the large memory bandwidth provided on this
architecture. It should be noted here that, although our comparison is not
an exhaustive benchmark, it does give an indication into the value of
each architecture for scientific computing, where the solution of
sparse linear systems is one of the most common operations. Since
sparse linear algebra has different hardware requirements than more
common benchmarks, such as High Performance Linpack (HPL), due to its
low computation-to-data-access ratio and irregular memory accesses
SpMV provides a valuable benchmarking alternative, as proposed by
Heroux and Dongarra~\cite{HerouxDongarra2013}.

\section{Summary and Discussion}
\label{sec:discussion}

In this paper we present an analysis of the scaling properties of
SpMV using a hybrid MPI/OpenMP extension to the PETSc library. We
compare hybrid vector-based and task-based algorithms with a pure-MPI
variant using large matrices generated by the open-source CFD code
Fluidity. We describe an extension to the traditional task-based
approach, where the load balance among threads is optimised a-priori
according to the number of non-zeros in each row.

The thread-balancing extension is shown to give superior performance
when scaled to large numbers of compute nodes on a Cray XE6 and on
moderate numbers of nodes of a Fujitsu PRIMEHPC FX10 system. On an IBM
BlueGene/Q system the optimised implementations is the only approach
capable of sustaining a parallel efficiency of $>90\%$ on up to 16384
cores. The algorithm achieves this by improving the memory bandwidth
utilisation within a given compute node and by hiding MPI
communication latency. This comes at the cost of increased memory
latency effects on small numbers of cores, since the algorithm creates
an imbalance in input vector elements per thread. However, once the
main resource limitation of the algorithm shifts to memory bandwidth
the thread-balancing approach can improve performance significantly.

Furthermore, the thread-balancing approach enhances one of the
fundamental advantages of hybrid programming: By reducing the number
of MPI processes the inherent load imbalance among processes is
reduced at the expense of load imbalance among threads. This is
desirable, however, since we can deal with the thread imbalance
explicitly by caching an optimised thread partitioning with the
matrix. As a result, this approach improves work load balance and
memory bandwidth utilisation at the compute node level in order to
increase overall performance.

\section*{Acknowledgements} 

The work presented here was funded by Fujitsu Laboratories of Europe
Ltd. and the European Commission in FP7 as part of the APOS-EU project
(grant agreement 277481). 

This work made use of the facilities of HECToR, the UK's national
high-performance computing service, which is provided by UoE HPCx Ltd
at the University of Edinburgh, Cray Inc and NAG Ltd, and funded by
the Office of Science and Technology through EPSRC's High End
Computing Programme.

We acknowledge use of Hartree Centre resources in this work. The STFC
Hartree Centre is a research collaboratory in association with IBM
providing High Performance Computing platforms funded by the UK's
investment in e-Infrastructure. The Centre aims to develop and
demonstrate next generation software, optimised to take advantage of
the move towards exa-scale computing.

\bibliographystyle{ieeetr}
\bibliography{bibliography} 

\end{document}